\documentclass[english,american,twocolumn]{revtex4}
\usepackage[T1]{fontenc}
\usepackage[latin9]{inputenc}
\usepackage{amsmath}
\usepackage{amssymb}
\usepackage{graphicx}

\makeatletter

\providecommand{\tabularnewline}{\\}
\newcommand{\lyxdot}{.}

\@ifundefined{textcolor}{}
{%
 \definecolor{BLACK}{gray}{0}
 \definecolor{WHITE}{gray}{1}
 \definecolor{RED}{rgb}{1,0,0}
 \definecolor{GREEN}{rgb}{0,1,0}
 \definecolor{BLUE}{rgb}{0,0,1}
 \definecolor{CYAN}{cmyk}{1,0,0,0}
 \definecolor{MAGENTA}{cmyk}{0,1,0,0}
 \definecolor{YELLOW}{cmyk}{0,0,1,0}
}

\makeatother

\usepackage{babel}
\begin{document}

\title{Brushes of semiflexible polymers in equilibrium and under flow in
super-hydrophobic regime}

\author{K. Speyer$^{*\dag}$, C. Pastorino$^{*\dag}$}
\begin{abstract}
We performed molecular dynamics simulations to study equilibrium and
flow properties of a liquid in a nano-channel with confining surfaces
coated with a layer of grafted semiflexible polymers. The coverage
spans a wide range of grafting densities from essentially isolated
chains to dense brushes. The end-grafted polymers were described by
a bead spring model with an harmonic potential to include the bond
stiffness of the chains. We varied the rigidity of the chains, from
fully flexible polymers to rigid rods, in which the configurational
entropy of the chains is negligible. The brush-liquid interaction
was tuned to obtain a super-hydrophobic channel, in which the liquid
did not penetrate the polymer brush, giving rise to a Cassie-Baxter
state. Equilibrium properties such us brush height and bending energy
were measured, varying the grafting density and the stiffness of the
polymers. We studied also the characteristics of the brush-liquid
interface and the morphology of the polymers chains supporting the
liquid for different bending rigidities. Non-equilibrium simulations
were performed, moving the walls of the channel in opposite directions
at constant speed, obtaining a Couette velocity profile in the bulk
liquid. The molecular degrees of freedom of the polymers were studied
as a function of the Weissenberg number. Also, the violation of the
no-slip boundary condition and the slip properties were analyzed as
a function of the shear rate, grafting density and bending stiffness.
At high grafting densities, a finite slip length independent of the
shear rate or bending constant was found, while at low grafting densities
a very interesting non-monotonic behaviour on the bending constant
is observed. 
\end{abstract}

\affiliation{$^{*}$Departamento de Física de la Materia Condensada, Centro Atómico
Constituyentes, CNEA, Av.Gral.~Paz 1499, 1650 Pcia.~de Buenos Aires,
Argentina}

\email{pastor@cnea.gov.ar}

\affiliation{$^{\dag}$ CONICET, Avenida Rivadavia 1917, C1033AAJ Buenos Aires,
Argentina}

\maketitle

\section{Introduction}

Polymer brushes are enormously versatile systems, whose properties
can be tailored chemically and physically\cite{Advincula_04,doi:10.1021/cr900045a,POLA:POLA26119}.
They were studied with great interest due to their technological significance,
biophysical importance and theoretical subtleties and complexity.
They find important applications in colloid stabilization, lubrication,
and developments where friction\cite{Klein_94}, adhesion and wetting
properties are important\cite{Advincula_04}. They are also used as
coatings in ``smart surfaces'' to fine tune reversibly some property
of an interface upon changes in external stimuli, for example, PH\cite{Stratakis_10},
temperature, solvent quality\cite{Peng_12,Chen201094}. Brushes are
at the forefront of recent developments, ranging from responsive bio-interphases,
controlled drug-delivery and release systems, thin films and particles,
which act as sensors of minute amounts of analytes\cite{Peng_12,POLA:POLA26119}.
The geometry and curvature of the grafted surface contribute also
to distinctive properties of brushes, which can be grown in planar
or cylindrical geometries, in the surface of nanoparticles or into
the backbone of macromolecules (bottle-brushes)\cite{Milchev_12b,Advincula_04}.
An important biophysical system of end-grafted semiflexible polymers
is the endothelial glycocalyx layer, present in the inner side of
vasculature in animals and plants\cite{Weinbaum_07}. It has a complex
structure formed by a matrix of macromolecular carbohydrates, consisting
of proteoglycans and glycoproteins, that coat the surface of endothelial
cells\cite{Weinbaum_07,Deng_2012}. It is known to have important
functions as modulator of permeability in the exchange of water, primary
molecular sieve of plasma proteins, mechanotransducer of fluid shear
stress and as regulator of red and white blood cells\cite{Weinbaum_07}.
From a mesoscopic point of view, this branched comb-like polymer brush,
can be modelled as composed of semiflexible homopolymer chains whose
beads have the size of the side-chains\cite{Deng_2012}. In this model,
the glycocalix is closely related to the geometry we study in this
work.

Polymer brushes were studied intensely in equilibrium and non-equilibrium
conditions\cite{Binder_11,Lee_14,Milchev_12b} by computer simulations
and theory\cite{Milner_91}. However, the vast majority of work was
devoted to fully flexible polymer chains. Computer simulation of coarse-grained
systems studied friction of bearing brushes \cite{Galuschko_10,Kreer2002,T.Kreer2001a,Grest99}
and the flow of simple liquids or polymer melts confined in brush-coated
polymer channels as function of grafting density, interface properties
and flow intensity\cite{Pastorino_06,Pastorino_09,Mueller_08b,Goicochea_14,Pastorino_14}.
Milchev and Binder studied the structure of brushes formed by semiflexible
chains by Monte Carlo and molecular dynamics simulations\cite{Milchev_12}.
They found an interesting phase transition of the system under compression,
which presents buckling at moderate pressures and bending at higher
pressures\cite{Milchev_13}. The first bonds of the grafted polymers,
in this reference, are directed perpendicularly to the grafting plane,
as it is the case in our work. Using soft potentials within the Dissipative
Particle Dynamics (DPD) simulation scheme various groups studied different
aspects of a polymer brush exposed to flow\cite{Deng_2012,Goicochea_14,Goujon_04}.
Deng \emph{et al.}\cite{Deng_2012}\emph{ }included bending rigidity
in their model to account for semiflexible polymers in the context
of glycocalyx and studied brush height and slip length for Couette
(shear-driven) and Poiseuille (pressure-driven) flows. Benetatos \emph{et
al. }explored the morphology and in-plane collapse of a grafted layer
of attractive semiflexible chains, analysing the bundling of neighboring
chains\cite{Benetatos_13}. Kim \emph{et al} studied the height behavior
of the brush layer as function of chain stiffness, shear rate and
grafting density by Brownian hydrodynamics, lattice Boltzmann simulations
and mean field theories\cite{Kim_09}. More recently, Römer and Fedosov\cite{Roemer_15}
developed a theoretical model and compared with DPD simulations, a
brush of stiff polymers under flow. They found a good agreement of
model and simulations for brush height, velocity profile and apparent
flow viscosity, for a wide range of shear rates and grafting densities.
Active semiflexible polymers (filaments) have been studied to analyse
efficency of pumping of liquid\cite{Kim_06} and methachronal waves
in bidimensional arrays of grafted chains\cite{Elgeti_13}. In these
works the focus is in the hydrodynamic coupling of the semiflexible
chains and its role in transport efficiency.

Relevant to the present work are also the super-hydrofobic surfaces.
In these, surface texture or structure at the micro and/or nanoscopic
level, is used to enhance the intrinsic hydrophobic chemistry of the
surface, to produce highly non-wetting surfaces. Superhydrophobic
surfaces typically exhibit very high water-repellency, which translates
in high contact angles $\Theta$ of droplets deposited over them ($\Theta>150^{o}$).
This property is desirable for an important range of technological
applications, such us self-cleaning surfaces, anti-fog, anti-corrosion,
icephobicity and fluidic drag reduction\cite{Roach_08,Zhang_12}.
Polymer brushes and coatings are also a way to obtain super-hydrophobic
surfaces\cite{Stratakis_10,Genzer_00}. In recent years, new methodologies
were developed to produce specific super-hydrophobic surfaces on small
and large scales, suitable for basic research and production, respectively.
Despite this, the underlying basic mechanisms of the super-hydrophobic
effects are under discussion, beyond the basics developed a long time
ago by Wenzel, Cassie and Baxter\cite{Wenzel_36,Cassie_44}.This includes,
for example, the role of roughness at multiple scales, which has also
importance in the development of bio-mimetic surfaces, whose biological
counterparts, such as the lotus leaf, are still under active research\cite{Bixler_12}.
Mimicking nature, is a promising route to the developing of multifunctional
super-hydrophobic surfaces. Lotus leaf's morphology, for example,
was used to produce self-cleaning films and iridescent super-hydrophobic
surfaces were developed to obtain anisotropic wettability by copying
the structure of butterfly wings\cite{Zhang_12}.

Experimental results confirm the efficiency of super-hydrophobic coatings
to reduce the drag of objects moving in water\cite{Shi_07,Choi_06}
or, equivalently, increase the flow rate of liquid flowing in channels\cite{Shirtcliffe_09,Truesdell_06}.
The friction reduction is attributed to the air filling of local structure
in the solid-gas composite interfaces, which gives rise to surface
slip\cite{Zhang_12,PhysRevLett.97.266101,:/content/aip/journal/pof2/16/5/10.1063/1.1669400,PhysRevLett.101.226101}.

In this work, we study in equilibrium and under flow, the interface
of a liquid and a layer of end-grafted semiflexible polymers for different
bending rigidities and as a function of grafting density. The interaction
between grafted chains and liquid is chosen such that the interface
is hydrophobic and compatible with super-hydrophobic behavior, for
a wide range of grafting densities. A typical snapshot of the system
in the superhydrophobic Cassie-Baxter state, for the stiffest polymers
considered in this work, can be observed in Fig. \ref{fig:System_Snapchot}.
We studied a small number of bending rigidities, ranging from fully
flexible to highly stiff polymers, such that the polymer resembles
a pillar, typical of superhydrophobic surfaces, and its configurational
entropy is minimal. This implies also statistical Kuhn segments much
smaller, of the order, or much higher than the contour length of the
chains. In section \ref{sec:Simulation-techniques} we present the
simulation technique, details of the model, and the way in which we
imposed flow in the system. Simulation results are presented in section
\ref{sec:Results}, devoting subsections \ref{sub:Static-properties}
and \ref{sub:Flow-properties} to equilibrium properties and behavior
under flow, respectively. We provide a final discussion and conclusions
in Section \ref{sec:Conclusions}.

\section{Model and Simulation techniques\label{sec:Simulation-techniques}}

The polymer chains are simulated by the widely utilized coarse-grained
beadspring Kremer-Grest model\cite{Kremer_90,Grest_86}. A finite
extensible nonlinear elastic (FENE) potential models the bond interactions
between neighboring beads of the same chain.
\begin{equation}
U_{FENE}=\begin{cases}
-\frac{1}{2}kR_{0}^{2}ln\left[1-\left(\frac{r}{R_{0}}\right)^{2}\right], & r<R_{0}\\
\infty, & r\geq R_{0}
\end{cases}\,,
\end{equation}

where $R_{0}=1.5\sigma$, the spring constant $k=30\varepsilon/\sigma^{2}$,
and $r=\left|\boldsymbol{r}_{i}-\boldsymbol{r}_{j}\right|$is the
distance between neighboring monomers. This potential simulates the
correct dynamics of polymers for a variety of thermodynamic conditions\cite{Kremer_90,Grest_86}.
All monomers of the $N=10$ polymer chains we simulated, interact
with each other through a truncated and shifted Lennard-Jonnes potential 

\begin{equation}
U_{INT}=\begin{cases}
U_{LJ}(r)-U_{LJ}(r_{C})\,, & r<r_{C}\\
0\,, & r\geq r_{C}
\end{cases}
\end{equation}
with $U_{LJ}$ is the standard Lennard-Jones potential:
\begin{equation}
U_{LJ}=4\varepsilon_{ab}\left[\left(\frac{\sigma}{r}\right)^{12}-\left(\frac{\sigma}{r}\right)^{6}\right]\,,
\end{equation}
where $\sigma=1$ defines the length unit. $r_{C}$ is the interaction
cut-off radius: if the distance between particles exceeds $r_{C}$,
this interaction is turned off. For the interaction between particles
belonging to the polymer, the cut-off radius was set to the minimum
of the Lennard-Jones potential $r_{C}=\sqrt[6]{2}\sigma\simeq1.12\sigma$,
giving a purely repulsive force which is interpreted as a polymer
in good solvent conditions. For the liquid-liquid and liquid-polymer
interactions, the cut-off radius was set to twice the minimum of the
potential $r_{C}=2\sqrt[6]{2}\sigma\simeq2.24\sigma$, to include
attractive interactions. This allows the formation of liquid and gas
phases for temperatures below the evaporating point. The interaction
strength $\varepsilon_{ab}$ was set to unity for the polymer-polymer
and liquid-liquid interactions $\varepsilon_{pp}=\varepsilon_{ll}\equiv\varepsilon=1$,
and defines the energy unit. The temperature unit is therefore $\varepsilon/k_{B}$.
For polymer-liquid interactions the parameter was set to $\varepsilon_{pl}=1/3\varepsilon$,
representing the chemical incompatibility between species. Rough surfaces
with similar interaction potential and $\varepsilon_{pl}$ parameter
have been studied previously\cite{Barrat_99,:/content/aip/journal/jcp/135/22/10.1063/1.3663965,PhysRevLett.96.206101,Leonforte_11},
resulting in highly hydrophobic substrates and yielding to contact
angles $\theta\sim120^{o}$ for droplets deposited over that surfaces. 

The rigidity of the polymers is imposed by a harmonic potential 
\begin{equation}
U_{b}(\boldsymbol{r}_{i},\boldsymbol{r}_{i+1},\boldsymbol{r}_{i+2})=\frac{1}{2}k_{b}\theta_{i,i+1,i+2}^{2}\,,
\end{equation}
where $\theta_{i,i+1,i+2}$ is the angle formed by two consecutive
bonds , and is measured from the polymer backbone. The bending constant
$k_{b}$ is one of the parameters varied in this work to explore different
persistence lengths. It is important to note that the sum of bending
forces over all the beads is null, so the forces applied on the free
end bead are transmitted, in average, to the grafted bead. To induce
a privileged orientation of the polymer chains perpendicular to the
substrate, the bending force calculations are performed adding a phantom
bead exactly below the fixed end. This additional bond generates a
force on the second bead, favouring an orientation angle $\theta_{1}=90^{o}$
with respect to the wall. 

The walls at $z=0$ and $z=D$ were modelled as impenetrable flat
surfaces, which interact with monomers via a purely repulsive integrated
Lennard-Jonnes potential of the form
\begin{equation}
U_{wall}(z)=\left|A\right|\left(\frac{\sigma}{r}\right)^{9}+A\left(\frac{\sigma}{r}\right)^{3}\,,
\end{equation}
with $A=3.2\varepsilon$\cite{Pastorino_09,Pastorino_14}.

The grafted end-beads of the 10-bead polymers are placed at a distance
$z=1.2\sigma$ from the substrate, and the lateral positions is randomly
chosen. The number of grafted sites is determined by the grafting
density. The initial positions of the liquid particles are uniformly
random distributed, and a force switch-on mechanism is used to relax
the system towards thermodynamic equilibrium and eliminate the overlap
between beads. Production runs of typically $10^{7}$steps are done
after this relaxation and $\begin{gathered}10^{6}\end{gathered}
$ time steps of thermalization. 

We used the dissipative particle dynamics (DPD) scheme\cite{Hoogerbrugge_92,Groot_97}
to perform the simulations at constant temperature. An important feature
of this thermostat is that it accounts for correct hydrodynamic behaviour\cite{Espanol_95b},
which is essential to obtain reliable results out of equilibrium\cite{Soddemann_03,Pastorino_15,Pastorino_07}.
The equation of motion for a particle reads:
\begin{equation}
m\frac{d^{2}\boldsymbol{r}_{i}}{dt^{2}}=\boldsymbol{F}_{i}+\boldsymbol{F}_{i}^{D}+\boldsymbol{F}_{i}^{R}\,,\label{eq:newton}
\end{equation}
where $\boldsymbol{F}_{i}$ represents the conservative forces, $\boldsymbol{F}_{i}^{D}$
and $\boldsymbol{F}_{i}^{R}$ are the dissipative and the random forces,
respectively. The mass $m$ is set to unity for all particles. Random
and dissipative forces are applied in pairs, in such a way that local
momentum is conserved. The form for these forces is 

\begin{equation}
\boldsymbol{F}_{i}^{D}=\sum_{j\neq i}\boldsymbol{F}_{ij}^{D},\,\boldsymbol{F}_{ij}^{D}=-\gamma\omega^{D}(r_{ij})(\hat{\boldsymbol{r}}_{ij}\text{·}\boldsymbol{v}_{ij})\hat{\boldsymbol{r}}_{ij}\,,
\end{equation}

\begin{equation}
\boldsymbol{F}_{i}^{R}=\sum_{j\neq i}\boldsymbol{F}_{ij}^{R}\,,\boldsymbol{F}_{ij}^{R}=\zeta\omega^{R}(r_{ij})\eta_{ij}\hat{\boldsymbol{r}}_{ij}\,,
\end{equation}
where the notation $\boldsymbol{a}_{ij}=\boldsymbol{a}_{i}-\boldsymbol{a}_{j}$
is used for vectors, $\gamma$ is the friction constant, $\zeta$
denotes the noise strength, $\omega^{D}$ and $\omega^{R}$ are weight
functions. These parameters and functions are not completely free
to choose in order to satisfy the fluctuation-dissipation theorem.
This constrain impose the conditions $\zeta^{2}=2k_{B}T\gamma$, and
$[\omega^{R}]^{2}=\omega^{D}$. The parameter $\gamma$ was set to
unity in all simulations, because this value allows to maintain constant
temperature and doesn't produce a significative over-damping of conservative
forces\cite{Pastorino_07}. The random variable $\eta_{ij}$ has zero
mean and second moment $\left\langle \eta_{ij}(t)\eta_{kl}(t')\right\rangle =\delta_{ik}\delta_{jl}\delta(t-t')$.
The usual choice of the weight functions for continuous forces was
made\cite{Soddemann_03,Pastorino_07}.
\begin{equation}
[\omega^{R}]^{2}=\omega^{D}=\begin{cases}
(1-r/r_{C})^{2}\,, & r<r_{C}\\
0\,, & r\geq r_{C}
\end{cases}\,,
\end{equation}
 where the cut-off radius is chosen equal to that that of Lennard-Jones
interaction:$r_{C}=2\sqrt[6]{2}\sigma\simeq2.24\sigma$.

We define the time units in terms of the Lennard-Jones parameters
and the mass $\tau\equiv\sigma\sqrt{m/\varepsilon}$. The Velocity
Verlet\cite{Frenkel2002} integration scheme was used to integrate
the equations of motion \ref{eq:newton}, with time steps in the range
$2\text{·}10^{-4}\tau\leq dt\leq5\text{·}10^{-4}\tau$. Under these
conditions, it was possible to perform equilibrium and non-equilibrium
simulations, maintaining a constant temperature over the whole system.
We tested this with temperature profiles across the channel, obtained
from the mean square velocity of the particles.

The temperature value $T=0.8\varepsilon/k_{B}$ was chosen to set
the liquid far enough from the condensation point and to have a relatively
low vapor density\cite{Mueller_00}. In thermal equilibrium, the liquid
and gas number densities are $\rho_{l}=0.69\sigma^{-3}$ and $\rho_{v}=0.03\sigma^{-3}$,
respectively. Several systems were studied, varying the number of
grafted chains per surface area and bending constant. In order to
compare quantitatively the behaviour of the different systems, it
is necessary to maintain some magnitude constant for all cases. We
chose to maintain the liquid density in the bulk equal in all studied
systems to provide a consistent thermodynamic condition. Therefore,
the number of liquid particles was changed so that in every equilibrated
system, the bulk liquid density satisfied $0.693\sigma^{-3}\leq\rho_{l}\leq0.697\sigma^{-3}$.
The low tolerance implemented is due to the fact that the normal pressure
is a rapidly varying function of $\rho_{l}$. This means that the
total number of liquid particles compatible with the range of allowed
densities may vary in about $\sim300$ , which constitutes roughly
3\% of the total number of liquid particles in a typical system.

\begin{figure}[t]
\centering{}\includegraphics[clip,width=0.53\columnwidth]{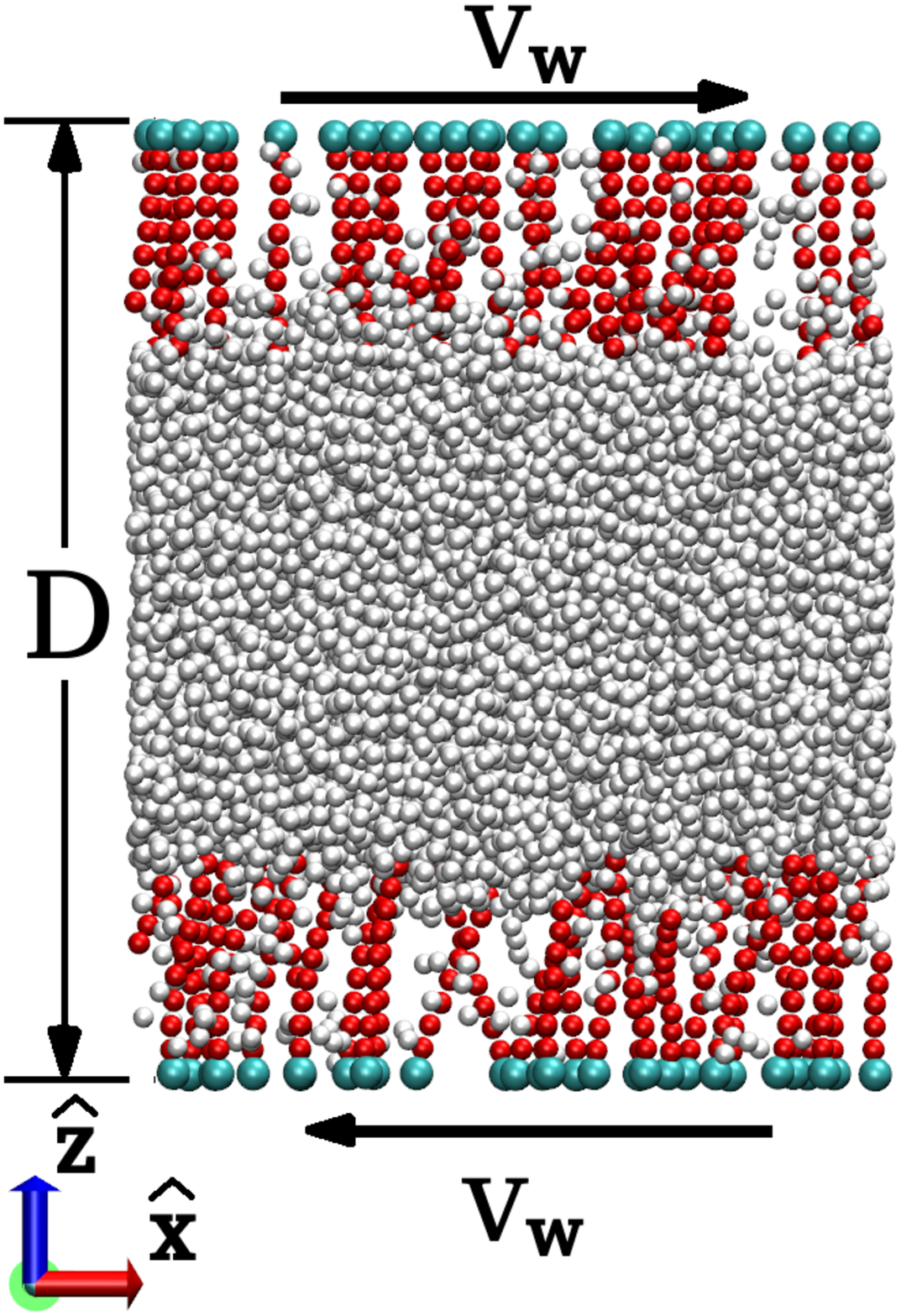}\includegraphics[clip,width=0.45\columnwidth]{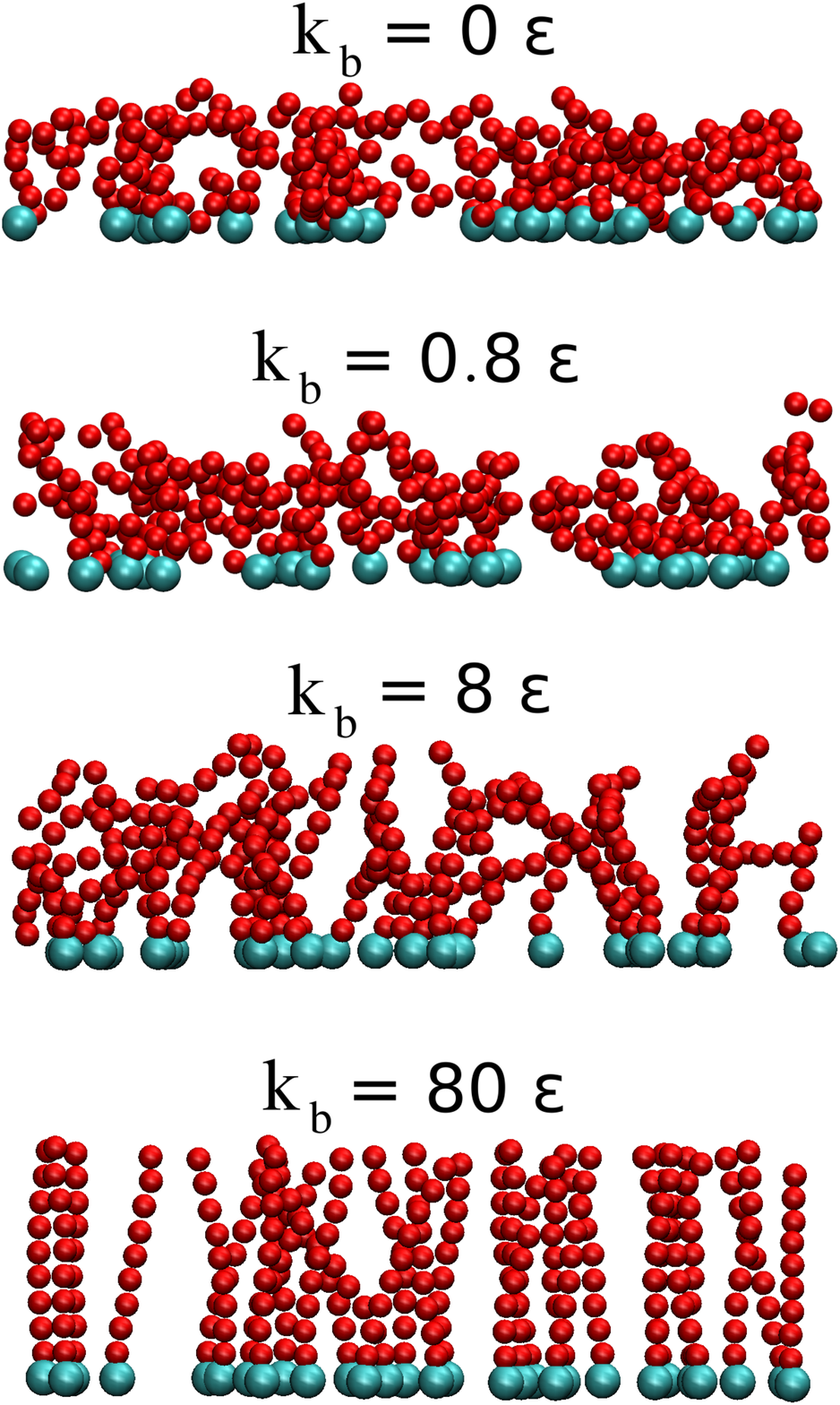}\protect\caption{\label{fig:System_Snapchot}Left panel: Snapshot of the simulated
system for a grafting density of $\rho_{g}=0.05\sigma^{-2}$ and bending
constant $k_{b}=80\varepsilon$ in equilibrium. Particles of the liquid
are shown in light gray, the polymer chains in the brush are presented
in red, while their en beads (steadily fixed to the wall) are shown
in light-blue. Right panel: Configuration of the polymer brushes for
the different bending rigidities studied (the liquid is not shown). }
\end{figure}

The grafting density $\rho_{g}$ is defined as the number of grafted
chains per unit area. Several systems with $\rho_{g}$ ranging from
$0.02\sigma^{-2}$ to $0.6\sigma^{-2}$ were studied. This values
of grafting density correspond to a mean distance between grafting
sites $l=7.1\sigma$ and $l=1.3\sigma$, respectively. Taking into
account that the contour length of the polymers is $l_{c}\backsimeq8.5\sigma$
($\begin{gathered}R_{end}\end{gathered}
\backsimeq4.5\sigma$ for fully flexible isolated chains), the range of $\rho_{g}$ selected
allows to study regimes from single chain condition, or the so-called
mushroom regime for flexible polymers, to a dense brush case. The
distance between walls is $D=40\sigma$ (see Fig. \ref{fig:System_Snapchot}),
and the lateral dimensions of the simulation box are $L_{x}=30\sigma$
and $L_{y}=20\sigma$.

Out of equilibrium simulations were produced by moving the walls at
constant velocity and in opposite directions. The polymers moving
with the wall drag the liquid and give place to a linear velocity
profile in the bulk of the liquid. The velocity of the walls was varied
from $v_{W}=0$ to $v_{W}=1.0\sigma/\tau$, which correspond to shear
rates between $\dot{\gamma}=0$ to $\dot{\gamma}=0.05\tau^{-1}$ and
Weissenberg number of $0\leq\begin{gathered}We\end{gathered}
\leq12$, according to the definition of the relaxation time of the polymers,
discussed below.

The relaxation times of the polymers for different bending constants
were estimated performing simulations without liquid and at low grafting
density ($\rho_{g}=0.01\sigma^{-2}$). Studying the autocorrelation
of the components of the end-to-end vector of the polymers it is possible
to extract a characteristic correlation time of the chains:
\begin{equation}
C_{i}(t)=\frac{\left\langle \left(R_{i}(0)-\left\langle R_{i}\right\rangle \right)\left(R_{i}(t)-\left\langle R_{i}\right\rangle \right)\right\rangle }{\left\langle \left(R_{i}-\left\langle R_{i}\right\rangle \right)^{2}\right\rangle }\,,
\end{equation}
where $R_{i}(t)$ is the i-component of the end-to-end vector. Four
values of the bending constant $k_{b}$ were studied over a wide range
of $k_{B}T$ values ($\begin{gathered}k_{b}/k_{B}T=0;\,1;\,10\end{gathered}
$ and $\begin{gathered}100\end{gathered}
$), giving the possibility to explore different rigidity regimes, from
fully flexible chains ($k_{b}=0\varepsilon)$, to very rigid rods
that resemble pillars ($k_{b}=80\varepsilon$). For $k_{b}=0\varepsilon,\,0.8\varepsilon\,,8\varepsilon$
the correlation follows an exponential decay and the relaxation time
can be obtained by fitting the curve $C(t)=exp(-t/\tau_{R})$. The
relaxation of the polymers with $k_{b}=80\varepsilon$ is more complicated
and does not follow a simple exponential decay. For short times, the
behaviour is exponential, but it presents in addition a longtime correlation
tail. This suggests complicated dynamics of the internal degrees of
freedom of the molecule, excited by thermal fluctuations. In order
to estimate a relaxation time in this case we fitted an exponential
law up to a correlation value of $C(t)=0.05$. This gives a relaxation
time which is roughly the double of the short-time correlation decay
$(\tau_{R}=125\tau$). 

Relaxation times of the grafted chains were also measured in simulations
including the liquid . The results differ in less than 30\% from those
measured without liquid, except for the case of $\begin{gathered}k_{b}=80\varepsilon\end{gathered}
$. For the stiffest chains ($\begin{gathered}k_{b}=80\varepsilon\end{gathered}
$), we found that the collisions between liquid particles and the free
end of the polymers decrease dramatically the relaxation time, turning
out in a lower value than that of fully flexible chains. Taking this
into account, it is evident that the relaxation times can be highly
dependent on the particular system where the polymers are embedded.
In order to obtain a characteristic polymer property, we decided to
use the relaxation times calculated in absence of liquid, which are
shown in Table \ref{tab:Persistence-length}. We do not intend to
study the complex dynamics of these highly rigid semiflexible polymers,
but to estimate an order of magnitude of their relaxation times. A
thorough study should be conducted to understand the relaxation kinetics
of the these semiflexible chains.

\begin{table}[t]
\centering{}%
\begin{tabular}{|c|c|c|c|c|}
\hline 
$k_{b}\,[\varepsilon${]} & $l_{p}/l_{c}$ & $l_{p}/\left\langle a\right\rangle $ & $k_{b}/k_{B}T$ & $\tau_{R}\,[\tau]$\tabularnewline
\hline 
\hline 
0 & 0.16 $\pm$ 0.01 & 1.4 $\pm$ 0.1 & 0 & 17\tabularnewline
\hline 
0.8 & 0.22 $\pm$ 0.03 & 2.0 $\pm$ 0.3 & 1 & 18\tabularnewline
\hline 
8 & 1.12 $\pm$ 0.06 & 10.1 $\pm$ 0.5 & 10 & 50\tabularnewline
\hline 
80 & 10.1 $\pm$ 1.7 & 91 $\pm$ 15 & 100 & 260\tabularnewline
\hline 
\end{tabular}\protect\caption{\label{tab:Persistence-length}Persistence length ($l_{p}$) over
contour length ($l_{c}$) for every studied bending constant ($k_{b}$)
studied. In the third column is presented the persistence length over
mean bond length ($\left\langle a\right\rangle $). The persistence
length covers a wide range of polymer rigidities, from fully flexible
chains, to highly rigid rods. The values were obtained from independent
simulations of isolated chains. The last column shows an estimate
of the relaxation time of the grafted polymers, obtained from the
correlation time of the end-to-end vector autocorrelation function.}
\end{table}
\vspace{250px}

In Table \ref{tab:Persistence-length} the values for the bending
constant are presented and compared to the energy of the typical thermal
fluctuations. It is also shown, the persistence length $l_{p}$ over
the contour length of the chains $l_{c}$. We point out that both
cases $l_{p}<l_{c}$ and $l_{p}>l_{c}$ are studied. For $k_{b}/k_{B}T\gg1$
the relation $l_{p}\simeq k_{b}/k_{B}T$ holds \cite{Milchev_13}.
To measure the persistence length, isolated chain simulations were
performed under the same temperature and using the same interaction
potentials as in the simulations for the brush-liquid system. The
segment correlation $cos(\theta_{\left|i-j\right|})=\left\langle \frac{\boldsymbol{a}_{i}\text{·}\boldsymbol{a}_{j}}{\left|\boldsymbol{a}_{i}\right|\text{·}\left|\boldsymbol{a}_{j}\right|}\right\rangle $
is measured for long simulations (over $10^{7}$ MD steps). Assuming
an exponential decay of the bond correlation with the bond number,
$\cos(\theta_{s})=\exp(-s\text{·}l_{p}/\left\langle \left|\boldsymbol{a}\right|\right\rangle )$,
it is possible to estimate the persistence length $l_{p}$. For the
stiffest cases $k_{b}=8\varepsilon$ and $k_{b}=80\varepsilon$, simulations
with larger polymers ($N=100$) were also carried out, to estimate
more precisely $l_{p}$, yielding similar results as those obtained
for $N=10$.

\section{Results\label{sec:Results}}

\subsection{Static properties\label{sub:Static-properties}}

\begin{figure}[t]
\begin{centering}
\includegraphics[width=0.98\columnwidth]{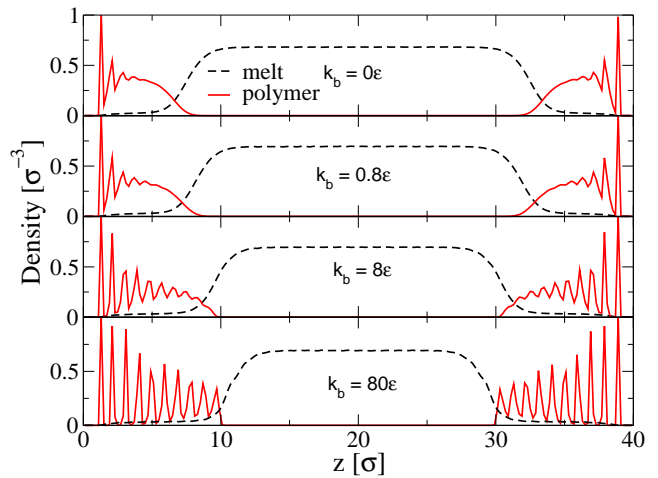}
\par\end{centering}

\protect\caption{\label{Density_prof}Density profiles for the liquid (dashed line)
and the brush layer (red, continuous line) for the different bending
constants. Upon increasing rigidity, the polymers stretch, and the
structure of the brush is enhanced. The grafting density is $0.2\sigma^{-2}$
for all the cases. }
\end{figure}

We begin our discussion by presenting the monomer density profiles
of the system in equilibrium for all the studied bending constants
$k_{b}$ (see Figure \ref{Density_prof}). Upon increasing the rigidity
of the bonds, the polymers stretch in the direction perpendicular
to the grafted surface. This builds up the structure of the brush,
noticeable in the sharp peaks of the polymer density profiles. The
chemical incompatibility between liquid and polymer particles, expressed
by the L-J parameters $\begin{gathered}\varepsilon\end{gathered}
_{pl}=1/3\begin{gathered}\varepsilon\end{gathered}
_{ll}=1/3\begin{gathered}\varepsilon\end{gathered}
_{pp}$, prevents the fluid from penetrating the brush in its liquid state.
The liquid density decreases rapidly in the vicinity of the polymers,
and only isolated particles enter in the brush, forming a gas in coexistence
with the liquid. This gives rise to a well-defined, narrow liquid-brush
interface, even for the lowest grafting density studied $\rho_{g}=0.02\sigma^{-2}$.
We note also that the fact the the liquid does not penetrate in the
the brush agrees with the definition of the Cassie-Baxter state, present
in nano-structured super-hydrophobic surfaces\cite{Cassie_44,Wang_12b,Tretyakov_13}.
This is illustrated for a relatively low grafting density $\rho_{g}=0.05\sigma^{-2}$
in Fig. \ref{fig:System_Snapchot}.

\begin{figure}[t]
\begin{centering}
\includegraphics[clip,width=0.98\columnwidth]{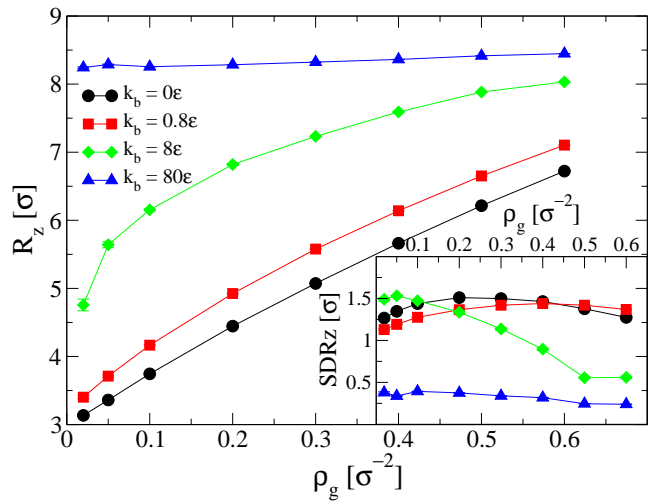}
\par\end{centering}

\protect\caption{\label{fig:Rz_vs_gd}Perpendicular component of the polymer end to
end vector $R_{z}$ as a function of the grafting density $\rho_{g}$
for the studied bending constants $k_{b}$. A linear relation may
be observed between the brush height and the number of chains per
surface area for the fully flexible case (black curve). The height
of the stiffest chains ($k_{b}=80$) is almost independent on the
grafting density. Inset: Standard deviation of perpendicular component
of the end-to-end vector, as a function of the grafting density. At
low $\rho_{g}$ the mobility of the semiflexible chains with $k_{b}=8\varepsilon$
is higher than the mobility of fully flexible chains.}
\end{figure}

The height of the polymer brush varies significantly upon increase
of the bending constant, due to the chain stretching. In Figure \ref{fig:Rz_vs_gd}
the vertical component of the end-to-end vector ($\boldsymbol{R}_{end}$)
is shown as a function of the grafting density for the different bending
constants $k_{b}$. $\boldsymbol{R}_{end}$ is defined as the vector
joining both end-beads of the same polymer. As it was observed for
the density profiles (Figure \ref{Density_prof}), increasing the
bending rigidity results in a thicker brush. The behaviour is similar
for all bending rigidities: the chains elongate in the $z$ direction
upon increasing $\rho_{g}$, due to excluded volume interactions,
as reported in Refs. \cite{Pastorino_06,Pastorino_09} for brushes
composed of fully flexible chains. This effect is less pronounced
for the more rigid chains ($k_{b}=8\varepsilon$) and barely noticeable
for the stiffest case ($k_{b}=80\varepsilon$), due to the finite
extension reachable by the chains. 

The inset of Figure \ref{fig:Rz_vs_gd} shows the standard deviation
of $\begin{gathered}R_{z}\end{gathered}
$, $SDR_{z}$ . This magnitude is associated with the typical chain
height variation, i.e. the mobility of the end bead in $z$ direction.
Interestingly, there is a non-monotonic behaviour for the most flexible
cases ($k_{b}=0\varepsilon$ and $k_{b}=0.8\varepsilon$ ). At low
$\rho_{g}$, $R_{z}$ increases with grafting density due to a larger
free space available for the chains. Naturally, this causes an increase
in $SDRz$. At high $\rho_{g}$, the presence of neighbour polymers
hinders the mobility of the end bead, decreasing $SDR_{z}$. The chains
with $k_{b}=8\varepsilon$ have the greatest mobility at low $\rho_{g}$,
but it decreases rapidly upon increasing $\rho_{g}$, because the
bending potential allows global buckling of the chains, but no local
bending. This means that the mobility of the end bead is highly dependent
on the mobility of the entire polymer. $SDR_{z}$ is fairly constant
as function of $\rho_{g}$ for the stiffest studied polymers ($k_{b}=80\varepsilon$
).

\begin{figure}[t]
\centering{}\includegraphics[clip,width=0.98\columnwidth]{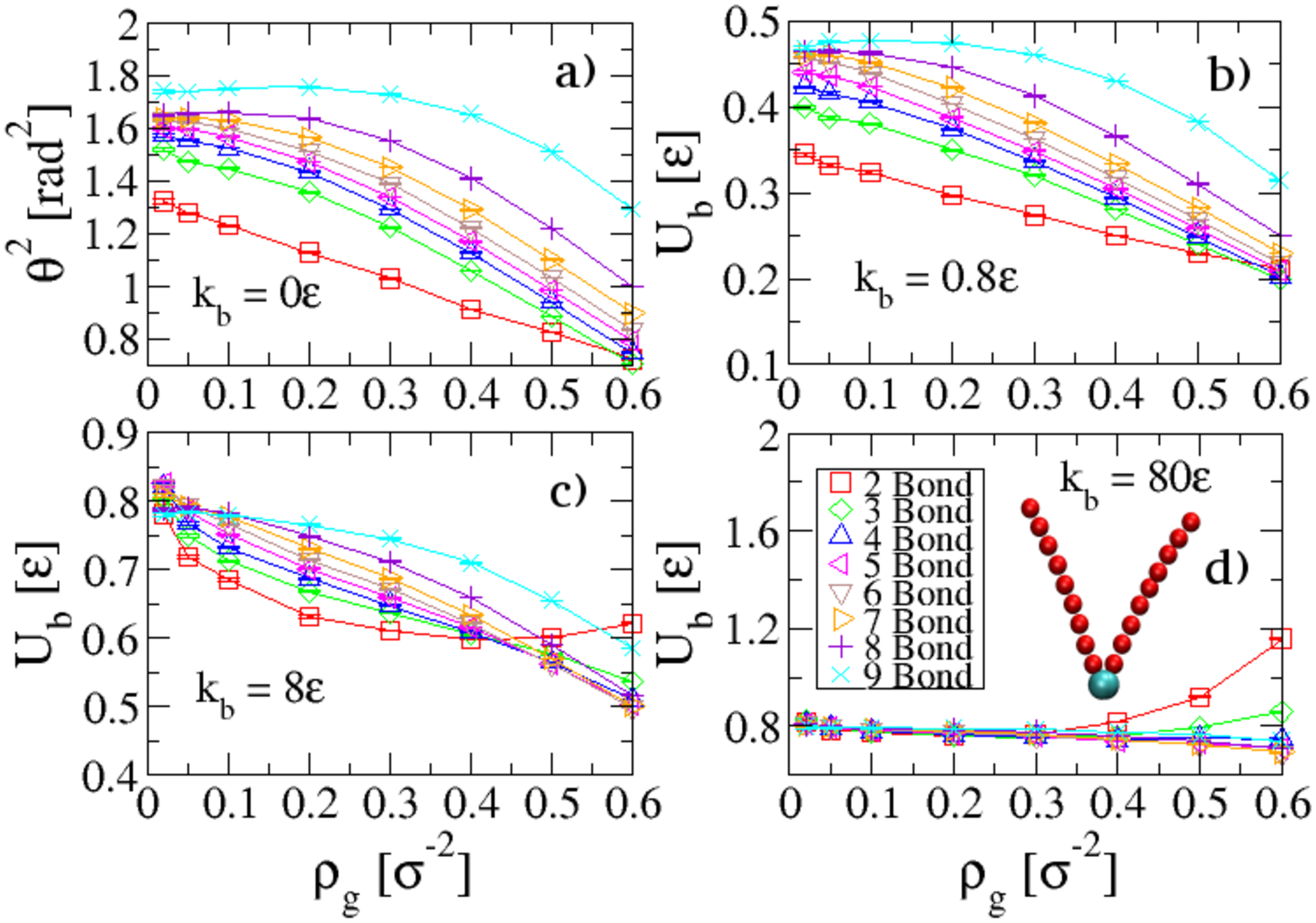}\protect\caption{\label{Ubond_vs_gd_vx0}Panel (a): mean square bond angle of individual
bonds versus grafting density ($\rho_{g}$) . Panels (b)-(d): Bond
energy ($U_{b}$) of individual bonds versus$\rho_{g}$. The first
bond is not shown due to its special behavior (see text). The bond
angles diminish upon increasing $\rho_{g}$, leading to an augment
in the chains end to end distance. The energy also increases with
the proximity of the bond to the free end bead. For $k_{B}=80$ the
bending energy per bond remains roughly the same for all grafting
densities.}
\end{figure}

To investigate the behavior of the internal degrees of freedom of
the polymers we measured the bond angles of each joint. Figure \ref{Ubond_vs_gd_vx0}
shows the mean bending energy per bond $U_{b}=1/2k_{b}\left\langle \theta_{i}^{2}\right\rangle $
as a function of $\rho_{g}$, where $\begin{gathered}\theta\end{gathered}
_{i}$ is the angle formed by two consecutive bonds in the i-th joint. Naturally,
$\theta$ decreases upon increasing the bending constant $k_{b}$,
but this does not translate immediately in increment of $U_{b}$.
At low $\rho_{g}$, increasing the bending constant from $k_{b}=0.8\varepsilon$
to $k_{b}=8\varepsilon$ roughly doubles the bending energy stored
in each bond, but increasing the bending constant from $k_{b}=8\varepsilon$
to $k_{b}=80\varepsilon$ varies $U_{b}$ in less than 5\%. For the
lowest grafting density, the increase in the end-to-end distance of
the polymers with persistence length $l_{p}$ is in good agreement
with the Worm-like chain or Kratky-Porod model\cite{rubinstein_colby}.

Upon increasing the grafting density, the excluded volume interactions
induce polymer stretching ( see Figure \ref{fig:Rz_vs_gd}), which
translates in a decrease in the bond angles. It can also be observed
that bond angles near the free end-bead tend to be larger than those
near the grafted end. Particularly, the last bead is only subject
to the bending force of the last bond and is constantly colliding
with liquid particles, which leads to a greater variance. 

The first \textquotedbl{}orientation\textquotedbl{} bond (not shown
in Figure \ref{Ubond_vs_gd_vx0}) behaves differently than the other
bond angles of the polymers, due to the constraints to which it is
subjected to. For low bending constants ($k_{b}=0\varepsilon$ and
$k_{b}=0.8\varepsilon$ ), the orientation angle is always lower than
the chain bond angles, because the wall limits the values this angle
can take. This reasoning is not true for the most rigid polymers ($k_{b}=8\varepsilon$
and $k_{b}=80\varepsilon$ ), because the strong bending forces bound
the internal bond angles as well as the orientation bond, and the
behaviour for both, internal and orientation angles, is similar at
low $\rho_{g}$. At $\rho_{g}\geq0.2\sigma^{-2}$ the presence of
overlapping grafting points force the polymers to take high orientation
angles as observed in the inset of Figure \ref{Ubond_vs_gd_vx0}.
This leads to a pronounced increase in the orientation bond energy,
which comes only from the way in which the sample is generated. The
grafting points are taken from a uniformly random distribution in
the x-y plane and, consequently, the probability of producing a sample
with overlapping grafting beads increases with grafting density.

\subsection{Flow properties\label{sub:Flow-properties}}

\begin{figure}[t]
\centering{}\includegraphics[clip,width=0.98\columnwidth]{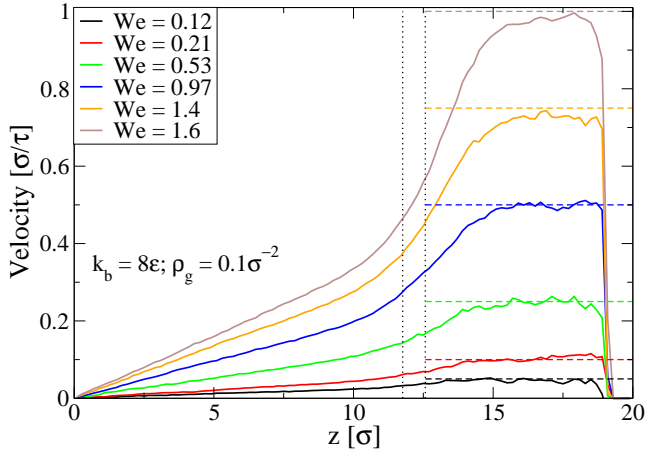}\protect\caption{\label{fig:Velocity-profiles-as}Velocity profiles as function of
shear rates for a grafting density $\rho_{g}=0.1\sigma^{-2}$ and
bending constant $k_{b}=8\varepsilon$. A linear profile can be observed
in the bulk of the channel. In the vicinity of the brush-liquid interface
(vertical dotted lines) there is a deviation from the linear dependence.
The noisy section of the profile near the wall, corresponds to the
fluid in its gaseous state. The wall velocity (and extension of the
brush layer) is indicated with dashed horizontal lines.}
\end{figure}

The system was taken out of equilibrium by moving the walls at constant
velocity in opposite directions\cite{Pastorino_06,Pastorino_14}.
This induces a flow with a linear velocity profile in the liquid phase
indicated in Figure \ref{fig:System_Snapchot}. In Figure \ref{fig:Velocity-profiles-as}
the symmetrized velocity profile of the liquid is shown for various
shear rates. In the bulk of the liquid, a linear dependence can be
clearly observed, as expected for a simple liquid under these boundary
conditions. In the vicinity of the interface (vertical dotted lines),
the behaviour changes and near the wall the free gas particles match
the wall velocity. It is interesting to note that the liquid layer
in contact with the polymer brush does not match the wall velocity,
i.e. the no-slip condition is violated and there is a partial slip
of the liquid on the brush. The shear rate $\dot{\gamma}$ was obtained
by fitting the linear velocity profile in the bulk of the liquid.
$\dot{\gamma}$ increases with the wall velocity, but also depends
on the grafting density $\rho_{g}$ and the bending constant $k_{b}$.
We define a Weissenberg number $We$, to quantify the shear rate with
the natural relaxation of the semiflexible polymers. $We$ is defined
as the shear rate ($\dot{\text{\ensuremath{\gamma}}}$) times the
relaxation time ($\tau_{R}$) of an isolated chain at the same temperature
than the simulated sample: $We=\dot{\text{\ensuremath{\gamma}}}\tau_{R}$.
As shown in table \ref{tab:Persistence-length}, the relaxation time
is highly dependent on the bending constant.

\begin{figure}[t]
\centering{}\includegraphics[width=0.98\columnwidth]{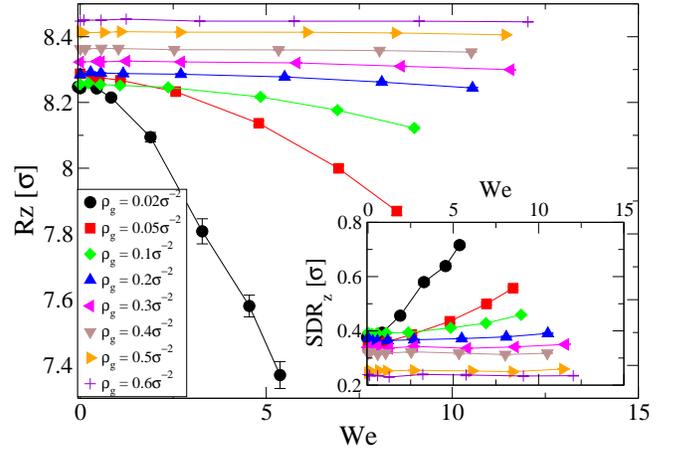}\protect\caption{\label{fig:Rz_vs_Shear}Perpendicular component of the end to en vector
$R_{z}$ for the polymer rods ($k_{b}=80\varepsilon$) as a function
of Weissenberg number for different grafting densities. For low $\rho_{g}$,
the stretched polymers lean in the direction of the flow and the mean
value of the normal component decreases. For high grafting the presence
of near neighbour chains hinders the leaning of the polymers and this
effect does not take place. Inset: Standard Deviation of $R_{z}$
as a function of Weissenberg number. Upon increasing the shear rate,
the mobility of the free end increases for low grafting densities.
The rapidly flowing particles of the liquid collide with the polymer,
bending it and increasing the standard deviation.}
\end{figure}

At high grafting densities, the density profiles are almost independent
of the shear velocity, while at low $\rho_{g}$, the brush height
diminishes upon increasing shear rate. In Figure \ref{fig:Rz_vs_Shear},
the mean vertical component of the end to end vector ($R_{z}$), for
the stiffest chains $k_{b}=80\varepsilon$, is presented versus Weissenberg
number $We$, for various grafting densities. A similar behaviour
for $R_{z}$ was observed as a function of $We$, for all the studied
values of bending constants. The collisions with the liquid particles
force the polymer chains to lean in the direction of the flow, thus
decreasing their height. For the rod-like polymers ($k_{b}=80\varepsilon$)
there is also an increase in the mobility of the end bead in the direction
perpendicular to the walls. To highlight this effect, the standard
deviation of $R_{z}$ is plotted as a function of $We$ in the inset
of Figure \ref{fig:Rz_vs_Shear}. For the lowest grafting density
($\rho_{g}=0.02\sigma^{-2}$), $SDR_{z}$ doubles it's value when
the shear rate is increased from $We=0$ to $We=10$. This effect
was not observed for other bending constants ($k_{b}=0\varepsilon,\,0.8\varepsilon,\,8\varepsilon$),
because the mobility is less hindered by the bending rigidity. This
data agrees with the model and simulations carried out by Römer and
Fedosov\cite{Roemer_15} in a similar system, but with hydrophilic
brushes. The shear rates studied in our work correspond to $\tilde{\dot{\gamma}}\equiv\dot{\gamma}\eta l_{c}^{3}/k_{B}T$
in the range $0\leq\tilde{\dot{\gamma}}\leq40$.

\begin{figure}[t]
\begin{centering}
\includegraphics[clip,width=0.98\columnwidth]{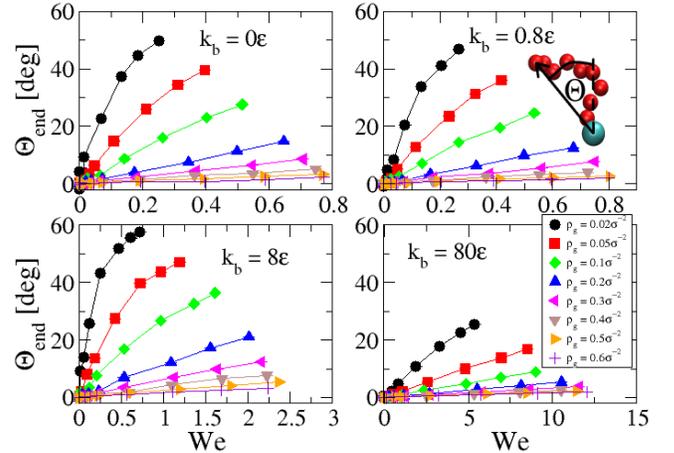}
\par\end{centering}

\protect\caption{\label{fig:Theta_vs_Shear}Mean end to end inclination angle of the
chain (defined in the inset), as a function of the Weissenber number
for the studied grafting densities and bending constants. For lower
grafting densities the polymers lean more, because there are less
excluded volume interactions with other neighbours. $\Theta_{{\rm end}}$
is greater for the chains with bending constant $k_{b}=8\varepsilon$
than for the more flexible ones. }
 
\end{figure}

In Figure \ref{fig:Theta_vs_Shear} the end-to-end inclination angle
(shown in the inset) is presented as a function of the Weissenberg
number $We$ for all the studied $k_{b}$ and $\rho_{g}$ . The inclination
angle is calculated as $\Theta_{end}=tan^{-1}(R_{z}/R_{x})$, where
$R_{x}$ and $R_{z}$ are the components of the end-to-end vector
in the shear direction and perpendicular to the walls, respectively.
$\Theta_{end}$ decreases upon increasing $\rho_{g}$, because excluded
volume interactions among beads of neighboring chains hinder chain
inclination. This is the same effect that causes the stretching of
fully flexible chains in equilibrium\cite{rubinstein_colby}. As expected,
increasing the shear rate induces a larger end-to-end angle, which
means that the chains tilt in the direction of the flow. For $\rho_{g}\geq0.2\sigma^{-2}$
the inclination angle is always lower than $20\text{º}$, and the
increment of $\begin{gathered}\Theta_{end}\end{gathered}
$ is linear with the shear rate. For lower grafting densities and high
$We$, there is a non-linear behaviour, due to saturation effects.
The dependence of $\Theta_{end}$ on the rigidity of the polymer is
more complex. There are two different regimes according to the relation
between persistence and contour lengths: a) $l_{p}<l_{c}$ and b)
$l_{p}>l_{c}$. If the persistence length is lower than the contour
length, increasing $k_{b}$ (rigidity parameter) produces a higher
end-to-end angle of the chains under shear. The consecutive bonds
tend to stretch coherently in the shear direction, due to the correlation
imposed by the bending potential. The pressure exerted by the simple
liquid on the polymers induce chain buckling and they take a banana-like
shape, maximizing $\Theta_{end}$ (see inset in Figure \ref{fig:Theta_vs_Shear}).
Increasing $k_{b}$ further (regime $l_{p}>l_{c}$) prevents the chain
to achieve high local bending angles, and the polymer stretches in
the vertical direction, disfavouring the buckling phenomenon. As seen
in Figure \ref{fig:Rz_vs_gd}, the stiffest chains $k_{b}=80\varepsilon$
in equilibrium are totally stretched vertically for all $\rho_{g}$.
It is important to recall that in this model there is a privileged
orientation angle $\Theta_{end}=0\text{º}$, because the first bond
$\theta_{1}$ is subjected to the same bending potential as the rest
of the polymer with an orientation perpendicular to the wall. 

\begin{figure}[t]
\centering{}\includegraphics[clip,width=0.98\columnwidth]{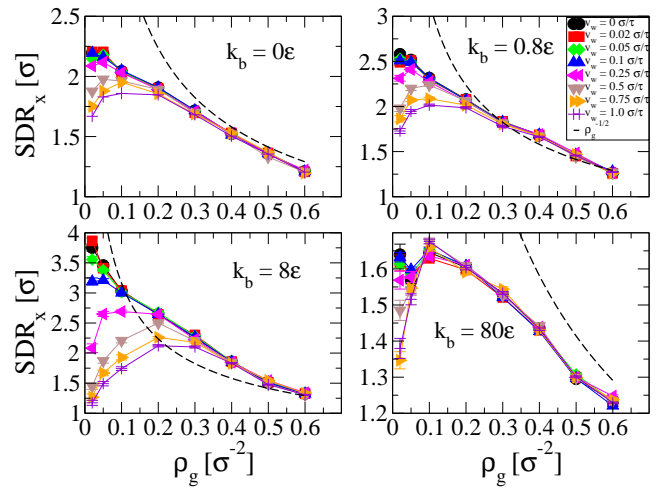}\protect\caption{\label{fig:SDRx_vs_gd}Standard deviation of the component parallel
to the flow direction of the end to end vector ($SDR_{x}$) as a function
of the grafting density ($\rho_{g}$) for the studied bending rigidities.
The mean distance between grafting points is shown with a black dashed
line. For low shear rates, the mobility of the polymers in the direction
of the flow decreases monotonically with the grafting density. For
high shear rates the behaviour of $SDRx$ is non monotonic with $\rho_{g}$.
At low grafting densities and large shear rates, the chains lean in
the direction of the flow and are not able to relax to the equilibrium
configuration, due to collisions with liquid particles. Upon increasing
the number of chains per surface area, there is a screening effect
that allows polymers to gain mobility in the shear direction. This
effect has a peak at intermediate $\rho_{g}$, decreasing then for
higher grafting due to the hindering of the mobility by excluded volume
interactions. }
\end{figure}

Another interesting quantity, is the standard deviation of the $\mathbf{R}_{end}$
($SDR_{x}$) component parallel to the shear direction. This magnitude
gives information about the typical mobility of the polymer's free
end-bead, around the mean value, in the shear direction. In Figure
\ref{fig:SDRx_vs_gd} is presented the dependence of $SDR_{x}$ on
$\rho_{g}$, for various shear rates and bending rigidities. It can
be observed that for $l_{p}<l_{c}$ ($k_{b}\lesssim8\varepsilon$),
increasing bending rigidity induces a higher displacement of the free
end. The pressure exerted by the liquid causes the more rigid polymers
($k_{b}=8\varepsilon$) to bend in a defined direction, due to correlation
between bonds, and thus enhances their displacement . Increasing $k_{b}$
above $l_{p}\eqsim l_{c}$ hinders the mobility of the chain, because
of the strong bending potential. This behaviour was previously observed
by Milchev and Binder\cite{Milchev_14b} for brushes in equilibrium
without explicit solvent. An interesting feature of $SDR_{x}$ is
that for high $\dot{\gamma}$, the dependence on the grafting density
is non-monotonic. At low $\rho_{g}$ the chains lean in the shear
direction and don't relax to the equilibrium configuration, thus having
a small mobility. Increasing the number of neighbouring chains produces
a screening effect, allowing the free end of the polymers to gain
mobility. $SDRx$ reaches a maximum at intermediate $\rho_{g}$, because
at high grafting densities the mobility of the free end decreases,
due to excluded volume interactions.

The typical displacement $SDR_{x}$ is compared to the mean distance
between grafting points (dashed line in Figure \ref{fig:SDRx_vs_gd}).
As expected for high $\rho_{g}$, the chain movement is limited by
the presence of neighbour polymers, and $SDR_{x}$ match the mean
distance between neighbours. At low grafting densities, the standard
deviation of $R_{x}$ is limited by the finite length of the polymer,
and by entropic effects. At intermediate $\rho_{g}$, $SDR_{x}$ is
higher than the mean distance between grafting points for semiflexible
chains ($k_{b}=0.8\varepsilon$ and $k_{b}=8\varepsilon$). It may
be also observed that for rigid rods ($k_{b}=80\varepsilon$) $SDRx$
is always less than the mean distance between grafting points. This
indicates that even for high $\rho_{g}$, the dynamics of this polymers
are greatly affected by the bending force.

We also checked the existence of cyclic motion dynamics in semiflexible
chains under shear, found in previous works for fully flexible isolated
grafted chains and polymer brushes\cite{Mueller_08b,Pastorino_14}.
We analysed the effect of stiffness in the mechanism of cyclic motion,
whose origin is attributed to spontaneous fluctuations of the polymer
chains towards regions of higher velocity in the liquid. According
to this description, a stiffer chain should have a reduction in the
cyclic motion, because the thermal fluctuations are reduced. This
is in fact what we observe. We computed the mean brush momentum velocity
profile $p(z)=\langle m\rho(z)v(z)\rangle$, exactly in the same way,
than previous references\cite{Pastorino_14} (not shown). The presence
of cyclic motion is evidenced when a positive $p(z)$ is observed
in the region in which the chains are directly exposed to and dragged
by the fluid and, at the same time, a negative $p(z)$ is observed
in an inner region, inside the brush. We found cyclic motion dynamics
for bending constants $k_{b}=0.8\varepsilon$ and $k_{b}=8\varepsilon$.
This was the case for low grafting density ($\rho_{g}=0.02\sigma^{-2}$
and $0.05\sigma^{-2}$), in which the dynamics of chains can be regarded
as independent, but also for higher grafting ($\rho_{g}=0.3\sigma^{-2}$)
already in the brush regime. The cyclic dynamics is indeed reduced
upon increase of the chain stiffness and vanishes completely for the
stiffest chains ($k_{b}=80\varepsilon)$. A more comprehensive characterization
of the effect and its consequences for flow inversion\cite{Mueller_08b,Pastorino_14}
could be an interesting future study.

\begin{figure}[t]
\begin{centering}
\includegraphics[clip,width=0.98\columnwidth]{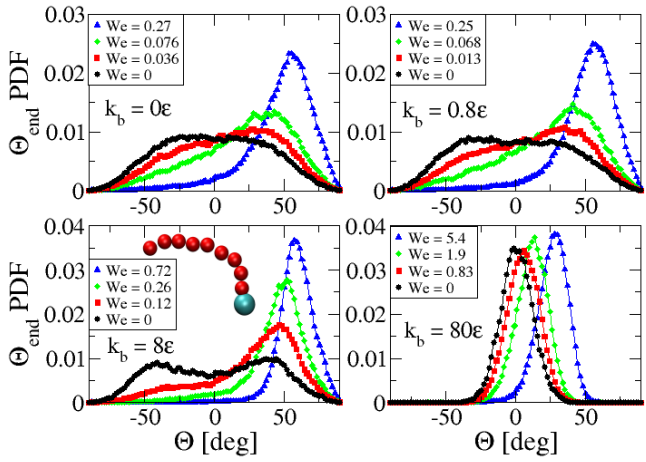}
\par\end{centering}

\protect\caption{\label{fig:ThetaPDF}Probability density function (PDF) of the end-to-end
angle of the polymers as a function of shear rate, for the lowest
grafting density $\rho_{g}=0.02\sigma^{-2}$. For $k_{b}=8\varepsilon$,
the equilibrium PDF is bimodal. The pressure exerted by the liquid
difficults the full stretching of polymers perpendicular to the wall,
giving rise to a non vanishing end-to-end angle. }
\end{figure}

In Figure \ref{fig:ThetaPDF} the probability density function (PDF)
of the end to end inclination angle $\Theta_{end}$ is presented for
various Weissenberg numbers and bending rigidities. The equilibrium
distribution has mean value $\left\langle \Theta_{end}\right\rangle =0$
for all cases, and high variance $var(\Theta_{end})=\left\langle \Theta_{end}^{2}\right\rangle -\left\langle \Theta_{end}\right\rangle ^{2}$.
Upon increasing the shear rate, the chains lean in the flow direction
and decrease their mobility, thus raising the mean value and reducing
the variance. The shape of PDF( $\Theta_{end}$) is highly non-symmetrical
with respect to the maximum for the samples under shear. This behaviour
is observable for all bending rigidities, except for $k_{b}=80\varepsilon$
that is less noticeable. It is interesting to note that all the PDF
deviate from a normal distribution, and particularly the tails follow
a non-gaussian decay. This non-gaussian behaviour was already found
in theory and experiments of single chains under shear.\cite{Winkler06} 

The buckling process for the case $k_{b}=8\varepsilon$ can also be
observed from the equilibrium PDF (black circles, in lower left panel
in Fig. \ref{fig:ThetaPDF}). There are two maxima, which correspond
to $\Theta_{end}=\pm40\text{º}$, and there is a local minimum at
$\Theta_{end}=0$, which corresponds to the free-end of the polymer,
verticaly aligned with the grafted monomer. The pressure exerted by
the liquid induces a buckling in the semiflexible chains , adopting
the shape shown in the inset of Fig. \ref{fig:ThetaPDF}. This effect
is most noticeable for chains whose persistence length is similar
to the contour length, due to the global order these polymers achieve.
For lower bending, a flat plateau at the maximum, more than a bimodal
distribution is observed. While for high bending rigidity ($k_{b}=80\varepsilon$),
the equilibrium distribution corresponds clearly to a normal curve.

\begin{figure}[t]
\centering{}\includegraphics[clip,width=0.98\columnwidth]{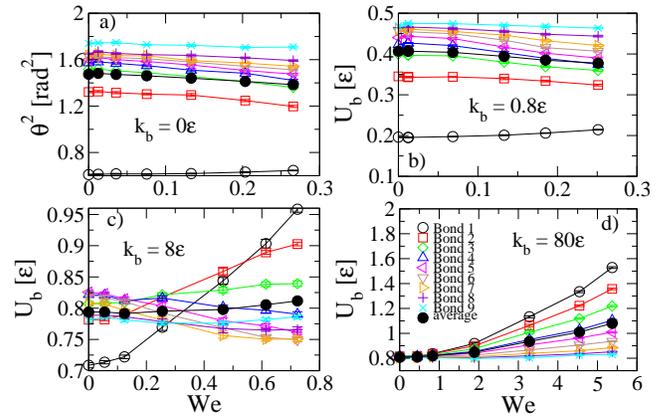}\protect\caption{\label{fig:ThetaBond_vs_Shear}Panel (a): Mean square angle of each
bond in the chain as function of Weissenberg number ($We$) for the
fully flexible case and $\rho_{g}=0.02\sigma^{-2}$. Panels (b)-(d):
Bond bending energy $U_{b}=1/2\text{·}k_{b}\text{·}\theta^{2}$ of
each bond as function of $We$ and $\rho_{g}=0.02\sigma^{-2}$. For
the most flexible polymers studied ($k_{b}=0$ and $k_{b}=0.8$),
the bonds in the chain decrease with increasing shear rate, and the
chains end-to-end distance increases. The orientation angle (Bond
1) increases because the chains lean on the direction of the flowing
liquid in the channel. The behaviour of the rods ($k_{b}=80$) is
different: all the bond angles increase upong rising the shear rate.}
\end{figure}

It's interesting to analyze how the internal degrees of freedom of
the polymers are affected by the shear stress. In Figure \ref{fig:ThetaBond_vs_Shear}
is presented the bending energy $U_{b}$ for each bond as a function
of the Weissenberg number $We$, for the lowest grafting density ($\rho_{g}=0.02\sigma^{-2}$).
Two different regimes can be distinguished: for $l_{p}\ll l_{c}$
(panels a and b) the orientation angle (Bond 1) is considerable smaller
than the internal angles, and the angles decrease upon increasing
shear rate, thus increasing the end to end distance. In this case,
the total bending energy is reduced with increasing shear rate. For
$l_{p}\gg l_{c}$ (panel d), the orientation angle is greater than
the internal angles, and the bending energy of all bonds increase
upon increasing $We$. It is also observed a decrease in the distance
between grafted and end beads. In this regime, the chains tend to
take a banana-like shape, leaning with the flowing liquid, and decreasing
the end to end distance. 

Polymers of $k_{b}=8\varepsilon$ are a limit case (panel c), because
the persistence length coincides with the contour length. The average
Bending energy (solid black dots in Fig. \ref{fig:ThetaBond_vs_Shear})
and the end-to-end distance (not shown) changes less than $2\%$.
In this limiting case, there are no important changes in the internal
structure of the polymers, which implies that the external perturbation
only alters the inclination angle. In addition, it can be observed
that the variation of bending energy between internal angles of the
same chain do not vary more than 30\% for $We\leq4$. This means that
the contribution of the internal angles to the total inclination of
the chain is roughly equidistributed along the polymer, as already
observed by Milchev and Binder\cite{Milchev_14b}.

\begin{figure}[t]
\begin{centering}
\includegraphics[clip,width=0.98\columnwidth]{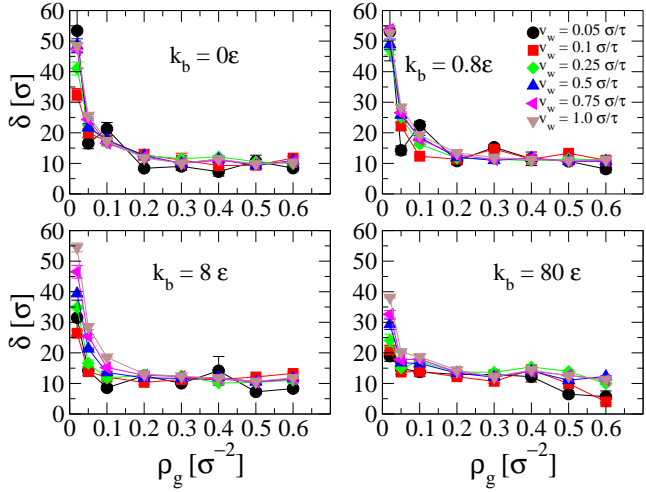} 
\par\end{centering}

\protect\caption{\label{fig:Slip-length vs gd}Slip length ($\delta$) as a function
of the grafting density ($\rho_{g}$) for the studied shear rates.
The slip length decreases rapidly upon increasing $\rho_{g}$ . For
intermediate and high grafting densities ($\rho_{g}\geq0.1\sigma^{-2}$)
$\delta$ takes roughly the same value $\delta\eqsim11\sigma$ for
every wall velocity and bending rigidity. All the studied cases present
non-zero slip length, indicating that the no-slip condition between
the polymer brush and the free melt does not hold. The rod-like polymers
($k_{b}=80\varepsilon$) are the ones that offer the most resistance
to the liquid flow for every grafting density and shear rate. }
\end{figure}

As mentioned above, the no-slip condition was violated in every system
studied. The friction reduction in systems with polymer brushes has
been a well known interesting feature for over a decade now\cite{Klein_94}.
We quantify here the effects of stiffness in the semiflexible polymers.
To quantify this phenomenon, the slip length ($\delta$) was measured
and plotted versus the grafting density $\rho_{g}$ in Figure \ref{fig:Slip-length vs gd}.
$\delta$ is defined as the distance at which the extrapolation of
the far field velocity profile of the liquid matches the substrate
velocity (see Ref. \cite{Tretyakov_13}). The liquid-polymer interface
position was defined as the $z$ coordinate at which the function
$f(z)=\rho_{l}(z)\text{·}\rho_{p}(z)$ presents a maximum, where $\rho_{l}$
is the liquid density, \foreignlanguage{english}{$\rho_{p}$} is the
polymer density and $z$ is the coordinate perpendicular to the walls.
At low $\rho_{g}$, the slip of liquid over the polymer brush is highly
dependent on the properties of the substrate. Varying the number of
grafted polymers per surface area and their rigidity yield different
results. In general, $\delta$ decreases rapidly upon increasing grafting
density, but the dependence with the stiffness ($k_{b}$) is non-monotonic.
At intermediate to high grafting densities ($\rho_{g}>0.1\sigma^{-2}$),
$\delta$ seems to saturate and takes a constant value, regardless
of the flexibility parameter $k_{b}$, the shear rate ($\dot{\gamma}$),
or the grafting density. Only for rigid rods ($k_{b}=80\varepsilon$),
there is a decrease in the slip at high grafting densities, and $We\lesssim1$.
We think that above $\rho_{g}=0.1\sigma^{-2}$, the slip length is
determined strongly by the brush-liquid compatibility which was fixed
in this study with the values $\varepsilon_{lp}=1/3\varepsilon_{ll}=1/3\varepsilon_{pp}$. 

\begin{figure}[t]
\centering{}\includegraphics[clip,width=0.98\columnwidth]{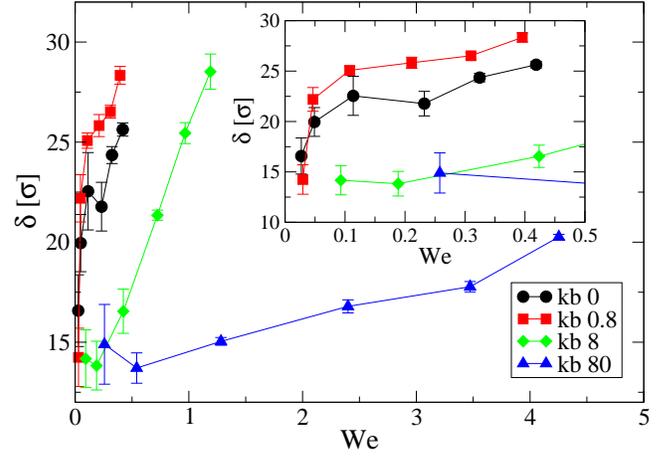}\protect\caption{\label{fig:SlipLength_vs_Shear}Slip length $\protect\begin{gathered}\delta\protect\end{gathered}
$ as a function of Weissenberg number $We$, for systems with low grafting
density ($\rho_{g}=0.05\sigma^{-2}$). The slip length is highly dependent
on the bending constant at lower grafting densities. The stiffest
chains ($k_{b}=80\varepsilon$) present less slip for all shear rates
studied. It is interesting to note the non-monotonic dependence of
$\delta$ with the bending constant $k_{b}$. Inset: detail of $\delta$
at low $We$. The slip is higher for the semiflexible polymers with
$k_{b}=0.8\varepsilon$, than for the fully flexible polymers.}
\end{figure}

In Figure \ref{fig:SlipLength_vs_Shear} is presented the slip length
$\delta$ as a function of the Weissenberg number $We$, for a low
grafting density $\rho_{g}=0.05\sigma^{-2}$. For every bending constant,
$\delta$ increases upon increasing shear rate, at low grafting densities,
as it was already reported for fully flexible chains\cite{Pastorino_06}
. A very interesting feature is that semiflexible chains with $k_{b}=0.8\varepsilon$
($l_{p}/l_{c}\simeq0.2$) induce higher slip length than fully flexible
chains, for all the studied shear rates. We think that a moderate
stiffness in the polymer could reduce the contacts of the chains with
the liquid as compared to the case of fully flexible chains in mushroom
regime. The dependence of $\delta$ on $k_{b}$ is non-monotonic.
Increasing further the stiffness (from $k_{b}=0.8\varepsilon$ to
$k_{b}=8\varepsilon$) decreases the slip length. Even though the
slip for $k_{b}=8\varepsilon$ is smaller than the slip of the more
flexible chains, the rate of change of $\delta$ with $We$ is roughly
the same for $We\geq0.5$. The rigid-rods brush ($k_{b}=80\varepsilon$)
is harder, and it shows a greater capacity to drag liquid along, than
the more flexible brushes. It has lower slip lengths in the whole
range of $\begin{gathered}We\end{gathered}
$ studied, and the growing slope is also lower. It is known that the
slip length is highly dependent on the normal pressure: the higher
the pressure, the smaller the slip\cite{Tretyakov_13}. In this case,
the normal pressure of the different systems varied less than 15\%,
and decreased monotonically with the bending constant. This implies
that the non-monotonic dependence of the slip length on the bending
constant can not be attributed only to variations in the normal pressure.

The force applied to the system in the direction of the wall displacement
as a function of the shear rate was also analyzed. The data from the
eight different grafting densities, four bending constants, and five
shear rate values collapse to a single straight line (not shown).
This means that the relation between the total force exerted on the
walls and the shear rate is independent of the polymer brush, and
only depends on the liquid. It was possible to extract the viscosity
of the simple liquid from the relation: $F_{x}=\eta\text{·}A\text{·}\dot{\gamma}$,
obtaining a value of $\eta=1.28\varepsilon\tau\sigma^{-3}$\cite{Pastorino_06,Goicochea_14}.
This essentially shows that the liquid behaves as a simple (newtonian)
liquid with a well defined viscosity for the whole range of shear
rates, and that the different morphologies of brush-liquid interface
or gas content inside the brush layer does not have a significative
role in the diffusion of momenta in the system.

\begin{figure}[t]
\centering{}\includegraphics[clip,width=0.98\columnwidth]{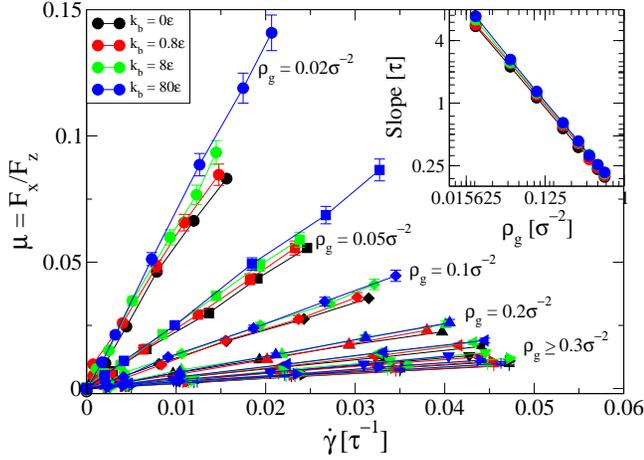}\protect\caption{\label{fig:Friction-coefficient-vs-shear}Friction coefficient $\mu$
of the system versus shear rate. Each group of curves corresponds
to a different grafting density $\rho_{g}$. It can be observed that
$\mu$ is proportional to the shear rate, and the slope is highly
dependent on the grafting density $\rho_{g}$. For systems with the
same $\rho_{g}$, the friction coefficient is higher, the higher the
bending constant . Inset: double-logarithmic scaling plot for the
slope of $\mu$ vs $\dot{\gamma}$, as a function of the grafting
density. The dependence of the slope on $\rho_{g}$ seems to follow
a power law.}
\end{figure}

Another useful quantity to compare different polymer brushes is the
kinetic friction coefficient ($\mu$), defined as the quotient between
shear and normal forces: $\mu=F_{x}/F_{z}$. Figure \ref{fig:Friction-coefficient-vs-shear}
presents $\mu$ versus $\dot{\gamma}$ as a function of the bending
constant $k_{b}$, for different grafting densities $\rho_{g}$. It
may be observed that the friction coefficient does not depend strongly
on the stiffness of polymer chains. The variation of $\begin{gathered}\mu\end{gathered}
$ between systems with the same grafting density and under the same
shear rate is in all cases less than $20\%$. Upon increasing the
grafting density, the friction coefficient decreases. . As encountered
in other studies \cite{Galuschko_10,Goicochea_14}, $\begin{gathered}\mu\end{gathered}
$ increases with shear rate. In this case it is observed that the relation
between $\mu$ and $\dot{\gamma}$ is linear for high $\begin{gathered}\rho_{g}\end{gathered}
$, because the normal force variation is very small with shear rate
for intermediate to high grafting densities. 

Every curve was fitted, performing a linear regression of the form
$\begin{gathered}\mu=m\dot{\gamma}\,,\end{gathered}
$ with only one parameter ($m$). The slope ($m$) of each curve is
plotted against $\rho_{g}$ in the inset of Figure \ref{fig:Friction-coefficient-vs-shear},
in a log-log scale. This implies that $m$ follows a power law of
the form $\begin{gathered}m=B(k_{b})\rho_{g}^{-1}\end{gathered}
$, where $\begin{gathered}B(k_{b})\end{gathered}
$ is a function that depends on $\begin{gathered}k_{b}\end{gathered}
$ only. The exponent of $\rho_{g}$ turns out to be $-1$ for all the
bending rigidities. Combining the last two equations is possible to
obtain a scaling law for the behaviour of the friction coefficient
$\mu$ as a function of the shear rate $\dot{\gamma}$ and the grafting
density $\rho_{g}$: 
\begin{equation}
\mu\thicksim B(k_{b})\dot{\gamma}\rho_{g}^{-1}
\end{equation}

It is interesting to note that the friction coefficient decreases
with increasing number of grafted chains per surface area. We think
that increasing $\begin{gathered}\rho_{g}\end{gathered}
$ hinders the liquid from penetrating the polymer brush. Adding polymer
chains in this superhydrophobic system could result in a smoother
liquid-brush interface, thus decreasing $\begin{gathered}\mu\end{gathered}
$. The components of the total force ($\begin{gathered}F_{x}\end{gathered}
$ and $\begin{gathered}F_{z}\end{gathered}
$) acting on the grafted end of the polymers have a different dependence
on the grafting density (not shown). While $\begin{gathered}F_{z}\end{gathered}
$ increases linearly with $\begin{gathered}\rho_{g}\end{gathered}
$, $\begin{gathered}F_{x}\end{gathered}
$ increases rapidly at low $\begin{gathered}\rho_{g}\end{gathered}
$, and at high grafting densities ($\begin{gathered}\rho_{g}>0.3\sigma^{-2}\end{gathered}
$) seems to reach a saturation value. We note however, that friction
phenomena are frequently scale dependent and friction coefficients
give different results in experimental measurements at nanoscopic
and macroscopic levels\cite{Nosonovsky_07}. They have usually contributions
from dissipative processes at various length scales. Also, macroscopic
concepts are not readily extrapolated down to small scale topographies\cite{Tretyakov_13}.

\section{Conclusions\label{sec:Conclusions}}

In this work, we studied comprehensively the properties in equilibrium
and under flow, of an interface of grafted semiflexible chains and
a simple liquid. We chose the interaction parameters between liquid
and polymer chains such that the liquid is in super-hydrophobic regime
on the top of the brush in the well-known Cassie-Baxter state. In
such, the liquid does not wet the interior of the grafted layer of
polymers and only molecules belonging to the coexisting gas phase
enter in the grafted polymer layer. The grafting density was varied
in a wide range, from mushroom regime of mostly isolated chains to
a dense brush with significant excluded volume interactions among
chains. Also the stiffness of the chains were studied in an ample
physical range, from fully-flexible chains to highly rigid chains,
in which the polymers resemble pillars, which were used as structures
to produce super-hydrophobic surfaces. We characterized the properties
of the chains by independent simulations in which we calculate the
persistence length, and the relaxation dynamics, by means of the time
correlation function of the end-to-end vector. 

We analysed the equilibrium structure of the grafted polymer layers
with density profiles, the behavior of the end-to-end vector components,
mean angles with the grafting plane, and bending energy of the chains.
The mean bending energy per bond behaves in a non-trivial way, while
for low bending constants ($k_{b}\sim k_{B}T$) the energy increases
with stiffness, for high $\begin{gathered}k_{b}\end{gathered}
$ it reaches a saturation value. We studied also the probability density
function of these angles in equilibrium and under flow, obtaining
non-gaussian distributions. We found a rich and distinctive behavior
of the chains for different bending rigidities. This is also reflected
in the typical shape that the chains adopt while supporting the liquid
layer on top of them. The most flexible polymers assume a typical
mushroom configuration for low grafting densities. For bending constants
corresponding to persistence lengths of the order of the chains' contour
length, the polymers bend in a banana-like shape to balance the pressure
exerted by the liquid. This leads to a higher displacement of the
end-to-end vector in the direction parallel to the substrate. In the
rigid rod limit, the polymers adopt typically a strait vertical configuration,
and buckling phenomenon is observed, due to the normal force exerted
by the liquid. By moving the confining walls at constant an opposite
velocities, we generate a linear Couette flow in the liquid and studied
the behavior of the brush layer under flow as function of grafting
density and shear rates. We measured the slip length of the liquid
and found a saturation to a well defined value upon increasing grafting
densities. At low grafting, where excluded volume effects are negligible
and the behavior is dominated by the properties of isolated chains,
we found that a moderate stiffness produces higher slip than fully
flexible chains. For even stiffer chains, the slip is reduced, giving
rise to a very interesting non-monotonic behavior of slip lengths
and velocities as a function of chain rigidity. We measured also the
friction coefficient of the interface by means of the mean normal
and shear forces, needed to impose a given shear rate. For a given
shear rate, we found that friction is inversely proportional to grafting
density. It could be interesting to expand these studies analysing
the differences in the reported behavior for liquids of lower chemical
incompatibility with the brush chains and the dynamics of droplets
on top of these polymer-coated surfaces of semiflexible chains. This
should be also of relevance for the fields of super-hydrophobic responsive
surfaces and microfluidics. 
\begin{acknowledgments}
Financial support through grants PICT-2011-1887, PIP 2011, INN-CNEA
2011, PICT-E 2014, is gratefully acknowledged. We thank also Marcus
Müller for fruitful discussions about different aspects of the present
work.
\end{acknowledgments}

%%% \bibliographystyle{rsc}
%%% \bibliography{database_1,bibtex}

\providecommand*{\mcitethebibliography}{\thebibliography}
\csname @ifundefined\endcsname{endmcitethebibliography}
{\let\endmcitethebibliography\endthebibliography}{}

\end{document}